# U.S. Technological Containment and the Rise of China's Open AI Ecosystem

Wang Jin[1†], Nadav Kunievsky[2†], Bowen Lou[3†], Tianshu Sun[4*], James Evans[2,5*]

**Affiliations:**

[1]Argyros College of Business and Economics, Chapman University; Orange, California, 92866, USA.
[2]Knowledge Lab, University of Chicago; Chicago, Illinois, 60637, USA.
[3]Marshall School of Business, University of Southern California; Los Angeles, California, 90089, USA.
[4]Cheung Kong Graduate School of Business; Beijing, 100006, China.
[5]Santa Fe Institute; Santa Fe, New Mexico, 87501, USA.

[†]These authors contributed equally to this work.
[*]Corresponding author. Email: tianshusun@ckgsb.edu.cn; jevans@uchicago.edu

Artificial intelligence (AI) has become a strategic domain of scientific leadership and geopolitical competition. Frontier AI systems increasingly depend on advanced semiconductors and large-scale compute (*1*), making access to technological infrastructure an increasingly important focus of policy. AI governance has become intertwined with geopolitical competition over semiconductors, compute, and technological standards (*2*), reinforcing the need for evidence-based policy amid intensifying global rivalry (*3*).

Over the past decade, the United States has pursued a dual strategy: accelerating domestic AI innovation while restricting China's access to frontier AI infrastructure. Export controls on advanced semiconductors and related technologies were intended to slow China's progress by limiting access to critical computational inputs and supply chains. Existing research has documented one important consequence of such restrictions: they appear to have spurred domestic technological substitution and greater self-reliance in China (*4*). However, technological substitution captures only part of the potential response to geopolitical constraint. Much less is known about whether technological containment is also associated with broader changes in how innovation is organized—through the models, software infrastructure, developer communities, and downstream actors involved in developing, improving, and diffusing new technologies. This question is particularly salient in AI, where open ecosystems increasingly connect model weights and architectures, advanced chips and compute services, open-source software, distributed developer and research communities, and downstream adaptation. Restrictions or uncertainty surrounding access to foreign-controlled technologies may increase the relative value of open, locally adaptable AI systems.

China's recent trajectory is consistent with this possibility. Over the past several years, Chinese policymakers and firms have promoted open-weight large language models (LLMs) and open-source infrastructure that support local deployment, modification, fine-tuning, and distributed development. This emphasis has become increasingly explicit at the highest levels of policy. At the opening of the 2026 World Artificial Intelligence Conference, Chinese President Xi Jinping called on China and the world to "Encourage open-source development and openness, cooperation and sharing, and comprehensively advance innovation in artificial intelligence science and technology". The emergence of DeepSeek and, more recently, Kimi K3 further illustrates the growing prominence of Chinese-origin open models despite persistent restrictions on advanced AI hardware and uncertainty surrounding access to frontier AI systems. These developments highlight the potential strategic value of openness under geopolitical constraint: beyond supporting innovation and ecosystem expansion, openness may strengthen technological resilience by reducing dependence on externally controlled technologies and expanding local capacity to adapt, improve, and deploy AI.

In this study, we trace the development of China's open AI ecosystem alongside tightening U.S. technological containment across several connected dimensions: government support for openness, the supply of open-weight models and open-source infrastructure, distributed participation and downstream improvement, and diffusion across developer, scientific, and commercial domains. Combining evidence from policy documents, open-model releases, GitHub activity, scientific publications, and patents, we document how these dimensions evolved over time. Before the escalation of export controls on high-performance chips, both the United States and China had adopted policies that advocated for open-source AI. Around major U.S. policy events, however, Chinese developers increased their engagement with open-source LLM repositories substantially more than U.S. developers, a pattern consistent with a broader ecosystem-level shift toward open

AI infrastructure under tightening technological constraints. We also observe increased research activity in areas tied to compute efficiency and lower-cost deployment. Chinese-origin open models subsequently diffused broadly through developer communities and gained substantial visibility in scientific research. Although such models remained largely absent from U.S. patent disclosures, U.S. company-affiliated researchers actively engage with them in open-access research, suggesting that their role in American commercial innovation may be undermeasured. Together, these findings reveal a broader consequence of technological containment that technological substitution alone does not capture. U.S. technological containment is associated not only with a redirection of Chinese AI activity, but also with the emergence of a more open, distributed, and potentially resilient ecosystem for developing, improving, and diffusing AI. This matters for policy because restrictions aimed at technological chokepoints may also shape the ecosystems that emerge in response. More broadly, AI competition concerns not only who controls frontier hardware and develops frontier models, but also whose ecosystems attract the researchers, developers, and organizations that build upon them.

## U.S. AI Policy and Technological Containment

China has emerged as the United States' principal competitor in AI (*5, 6*). Its rise reflects a broad national effort encompassing state-backed industrial policy, leading technology firms, a large scientific workforce, abundant data, and a vast domestic market (*7, 8*). Against this backdrop, U.S. policymakers sought both to strengthen domestic AI capabilities and to limit rivals' access to frontier AI infrastructure (*9-13*). AI competition has consequently expanded beyond models and capabilities to the technological chokepoints that enable frontier development.

Domestically, the United States has sought to strengthen the foundations of AI leadership through investments in research, talent, compute, standards, and public–private coordination. Initiatives including the American AI Initiative and the CHIPS and Science Act reflected this effort to expand domestic AI capacity while sustaining innovation and commercialization. At the same time, U.S. foreign policy increasingly identified China as the central strategic competitor in AI (*5, 6*). Policy attention focused particularly on advanced semiconductors, compute infrastructure, and the supply chains required for frontier AI development. Export controls became a principal instrument of this strategy (*12-15*). Beginning with firm-specific restrictions and expanding into broader semiconductor controls, the United States sought to limit China's access to advanced chips, chipmaking equipment, and related technical support. The October 2022 Bureau of Industry and Security controls marked a decisive shift by directly targeting the compute infrastructure underlying frontier AI systems (*12*), followed by further restrictions in 2023 and 2024 (*13-15*).

These policies raised the costs and uncertainty of AI development in China by restricting access to the most advanced chips and complicating the procurement of high-performance compute. The disruption is visible in the market for China-oriented accelerators, where U.S. licensing requirements on NVIDIA's H20 chips generated a multibillion-dollar inventory charge (*16*). Chinese model development reveals a corresponding adjustment: DeepSeek's V3 and R1 systems relied on NVIDIA H800, export-compliant accelerators developed for the Chinese market following earlier U.S. restrictions. These developments illustrate pressure associated with U.S. technological containment and China's capacity to innovate under constraint through export-compliant hardware, compute-efficient models, and software optimization. Tighter constraints on frontier infrastructure and uncertainty around proprietary model access may have increased the relative value of AI systems that can be locally deployed, adapted, and improved.

## The Rise of China's Open AI Ecosystem

China's response to growing technological constraints has extended beyond import substitution and support for proprietary national champions. An increasingly prominent element has been the development of an open AI ecosystem that broadens participation, accelerates downstream adoption, and reduces dependence on foreign-controlled technologies. Under geopolitical uncertainty, openness may become strategically attractive not despite technological competition, but partly because of it.

This orientation is visible in both policy and firm activity. China's 14th Five-Year Plan elevated open source into a national digital strategy by encouraging open-source communities, stronger governance mechanisms, and broader sharing of software, hardware designs, datasets, and application services (*7*). Because China's Five-Year Plans shape industrial priorities, public investment, and local implementation, these initiatives represent more than symbolic support for openness; they embed open-source development within a broader national technology strategy. More recent initiatives, including the 2025 Global AI Governance Action Plan *(17)*, further emphasized cross-border open-source collaboration, shared AI resources, and lower barriers to AI innovation and deployment. Chinese firms increasingly reflect this orientation. Alibaba released more than 100 open-weight models from its Qwen 2.5 family in 2024, DeepSeek's open-weight releases contributed to the growing prominence of Chinese-origin open models, and successful start-ups including Z.ai and Moonshot (producer of the popular Kimi models) shifted from hybrid to more open strategies. These developments rest on a rapidly expanding scientific base: China's funded AI and machine-learning publications, as well as its citation-weighted AI output, increased sharply relative to the United States (Fig. S1), helping position China at the forefront of global AI research and making it the world leader in AI publication volume (Fig. S2).

The strategic appeal of openness also reflects the changing economics and organization of AI development. As competition expands from frontier model development toward large-scale inference and application deployment, the ability to run, fine-tune, compress, and adapt models across diverse environments becomes increasingly important. More broadly, research on open innovation shows that firms compete not only through products, but also through ecosystems, standards, and developer participation (*18, 19*). Open platforms can mobilize distributed contributors, attract downstream complementors, and accelerate experimentation and adaptation. Android provides an earlier example of how openness can coordinate a broad ecosystem around a common technological platform. A related dynamic emerged with Huawei's OpenHarmony: amid restrictions on Huawei's access to parts of the Android ecosystem, the development of an open operating system reduced reliance on an externally governed platform while creating infrastructure around which other developers and firms could build. These advantages increasingly apply to open-weight LLMs. When access to proprietary systems or frontier infrastructure becomes more uncertain, open-weight and locally adaptable models may become relatively more valuable because they can be deployed, modified, and improved by distributed actors. China's turn toward open AI should therefore be understood not simply as greater support for openness, but as the emergence of a broader ecosystem that is more distributed, adaptable, and potentially more resilient under geopolitical constraint.

## Openness as Resilience Under Geopolitical Constraint

Technological containment can alter the relative value of different ways of developing and deploying AI. When access to frontier compute or foreign-controlled platforms becomes more uncertain, systems that can be independently deployed, modified, and broadly distributed become more strategically valuable. Our analysis of Chinese policy documents shows that openness is consistently linked to three related objectives. The first is ecosystem expansion: the 2015 "Internet Plus" strategy (*20*) encouraged open and collaborative innovation, with open release lowering barriers to experimentation and mobilizing developers, firms, and research institutions around shared infrastructure. A decade later, the 2025 "AI Plus" Opinions (*21*) explicitly called for promoting "a thriving open-source ecosystem." The second is governance and standards coordination: Chinese policies connect open AI to technical standards, interoperable interfaces, compliance frameworks, and intellectual property rules intended to support large-scale adoption (*7, 17*). The third is technological resilience: the 13th and 14th Five-Year Plans (*7, 22*) identify technological self-reliance as a national priority. Open-weight models and open-source infrastructure can support this objective by enabling local deployment and adaptation while reducing reliance on foreign-controlled technologies and centralized proprietary systems.

This strategic rationale is reinforced by the economic and technical advantages of open-weight LLMs. Such models can increase customizability, competition, transparency, and the distribution of technical capability across firms and developers (*23*). Recent evidence further suggests that open-weight models have helped push capabilities up while pushing costs down: the Stanford AI Index (*24*) reports that the inference cost of a system performing at the level of GPT-3.5 fell more than 280-fold between November 2022 and October 2024, while the performance gap between open-weight and closed models narrowed from 8% to 1.7% on some benchmarks. Open-source software infrastructure further lowers deployment barriers. For example, vLLM, an open-source LLM serving engine, uses its PagedAttention system to increase serving throughput by two to four times at comparable latency, facilitating local and lower-cost deployment.

Open weights also enable distributed downstream improvement. Developers can fine-tune released models for specialized tasks, allowing capabilities to continue evolving after the initial release. Task-specific fine-tuning can substantially improve performance relative to base models (*25*). Fig. S3 to Fig. S6 document this process directly, tracking how official model releases compare with their fine-tuned descendants across a range of standard reasoning and knowledge benchmarks. Within every size class, the best descendants repeatedly match or surpass their corresponding official releases, so that effective open-weight capability is set by the evolving community frontier rather than by the initial release alone. This descendant advantage is largest for small models and narrows as model size grows, consistent with large official releases operating closer to the attainable frontier.

These features yield a testable implication. If open AI serves partly as resilience infrastructure, policy events that increase uncertainty around access to frontier AI inputs should be associated with greater engagement with open and locally deployable systems. Fig. 1A shows that Chinese-origin open models increased rapidly after 2022 alongside growing policy emphasis on open-source communities, standards, interoperability, and platform coordination. Fig. 1B provides complementary behavioral evidence: forking of LLM-related repositories increased sharply around major U.S. policy events, with a substantially larger response among Chinese developers than among U.S. developers. Across the pooled shock window, mean weekly forks for China-associated LLM repositories increased by 0.143 forks per repository-week, compared with 0.012 for U.S.-associated repositories; the China-associated rise is 0.131 forks per repository-week larger. The divergence was most pronounced around export-control episodes that directly constrained access to advanced compute. This pattern is consistent with a policy-linked increase in the adoption of open, locally deployable AI infrastructure. A parallel shift is visible in research activity. Following the 2022 U.S. export controls, China's relative research emphasis increased in areas associated with compute efficiency and distributed deployment, including parameter-efficient fine-tuning, inference optimization, compression, and edge AI (Fig. 1C). These trends are consistent with technological constraints being associated not only with greater engagement with open AI infrastructure, but also with increased attention to techniques that make AI development less dependent on scarce frontier compute.

**Uneven Diffusion Across Institutional Domains**

The expansion of China's open AI ecosystem increased the availability of Chinese-origin open models and the infrastructure supporting their adoption. These models diffused broadly across decentralized domains, particularly developer communities and scientific research. Yet the extent and visibility varied across institutional settings.

Developer communities exhibited especially rapid uptake. Fig. 2C, together with Fig. S7 and Fig. S8, traces the diffusion of major open LLM families and shows sustained adoption among both Chinese and U.S. developers. Chinese-origin models such as Qwen and DeepSeek diffused rapidly through developer communities, while Chinese developers also extensively adopted U.S.-origin open-source models. This cross-border adoption highlights how open AI ecosystems can facilitate experimentation and recombination across national boundaries, even amid intensifying geopolitical competition. Although developers in both countries adopted these models at similar rates immediately following release, adoption continued to expand among Chinese developers while growth plateaued earlier among U.S. developers, producing substantially higher cumulative adoption in China.

Diffusion was similarly broad in scientific research. As shown in Fig. 2A, Chinese and U.S. researchers engage with Chinese-origin open models at broadly comparable rates. Patterns of engagement do not differ substantially between U.S.- and China-dominated institutions, suggesting that scientific communities continue to operate through relatively decentralized norms of experimentation and knowledge exchange, with model engagement driven more by utility and accessibility than by national origin.

A similar pattern appears in company-affiliated research. Fig. 2B traces references to Chinese- and U.S.-origin LLMs in arXiv publications affiliated with companies in both countries. Chinese and U.S. firms engage with models originating in both countries, indicating broad cross-national visibility in publicly disclosed research. Fig. S9 disaggregates these patterns by company origin and reveals modest home-country preference: Chinese firms reference Chinese-origin models somewhat more frequently, while U.S. firms display a comparable preference for U.S.-origin models.

The pattern changes markedly in the formal commercialization channel. We analyze references to LLMs in U.S. patent applications and granted patents. In contrast to scientific publications, the U.S. patent system displays a pronounced asymmetry. Patents frequently reference U.S.-origin and proprietary models such as LLaMA, ChatGPT, and Claude, yet references to Chinese-origin models such as Qwen and DeepSeek are nearly absent (Fig. 2B). Patent documents routinely reference external technologies and AI systems when doing so is useful for disclosure and uncontroversial. The selective absence of Chinese-origin models thus suggests that diffusion becomes increasingly filtered as innovation moves into institutionalized commercialization settings.

The divergence reflects differences in institutional context. Developer communities and scientific preprints operate through relatively decentralized norms that facilitate rapid experimentation, recombination, and distributed participation. Patent systems, by contrast, are embedded within legal review, compliance requirements, procurement considerations, and strategic disclosure incentives. In these settings, geopolitical exposure and regulatory scrutiny can shape which technologies organizations are willing to acknowledge and formally disclose. The near absence of Chinese-origin models from U.S. patents likey substantially understates these models' role in U.S. commercial innovation, particularly given their widespread and growing engagement by company-affiliated research and development.

## Discussion

Global competition in AI is increasingly a contest over ecosystems, not only frontier models and hardware. U.S. policy has focused heavily on controlling critical technological chokepoints, particularly advanced semiconductors and compute infrastructure. Yet our findings point to a broader consequence: constraints on frontier infrastructure may increase the strategic value of open, locally adaptable AI systems and encourage the development of alternative ecosystems around them. Technological containment may shape not only the pace of AI development, but also the ecosystems through which AI is developed, improved, and diffused.

This has important implications for scientific and commercial leadership. In science, open models lower barriers to experimentation, enable distributed improvement, and diffuse rapidly across research communities. The substantial presence of Chinese-origin models in U.S. scientific and company-affiliated research suggests that leadership depends not only on producing frontier models and hardware, but also on shaping the models, tools, and infrastructure on which global research builds. Commercial leadership follows a different path. The near absence of Chinese-origin open models from U.S. patent disclosures suggests that legal review, compliance requirements, procurement considerations, and strategic disclosure incentives sharply reduce visible adoption. Policy analysis should therefore connect scientific diffusion and commercial integration: openness expands access and use, but selective disclosure likely undercounts the influence of Chinese-origin open models on U.S. and global commercial innovation.

The strategic importance of openness is increasingly recognized in policy and industry. Recently, NVIDIA and other technology companies have argued that American AI leadership will depend not only on frontier models and hardware, but also on widely adopted open models that mobilize developers and strengthen the broader ecosystem (*26*). The U.S. Department of Energy's Genesis Open Models Initiative similarly reflects growing interest in open-weight models as infrastructure for scientific discovery and research (*27*). Sustaining AI leadership may thus require more than protecting technological chokepoints; it may also require cultivating open models, developer communities, research infrastructure, and credible pathways for commercial adoption. The future of AI leadership may hinge not only on who controls the technological frontier, but also on whose ecosystems become the foundations on which others build.

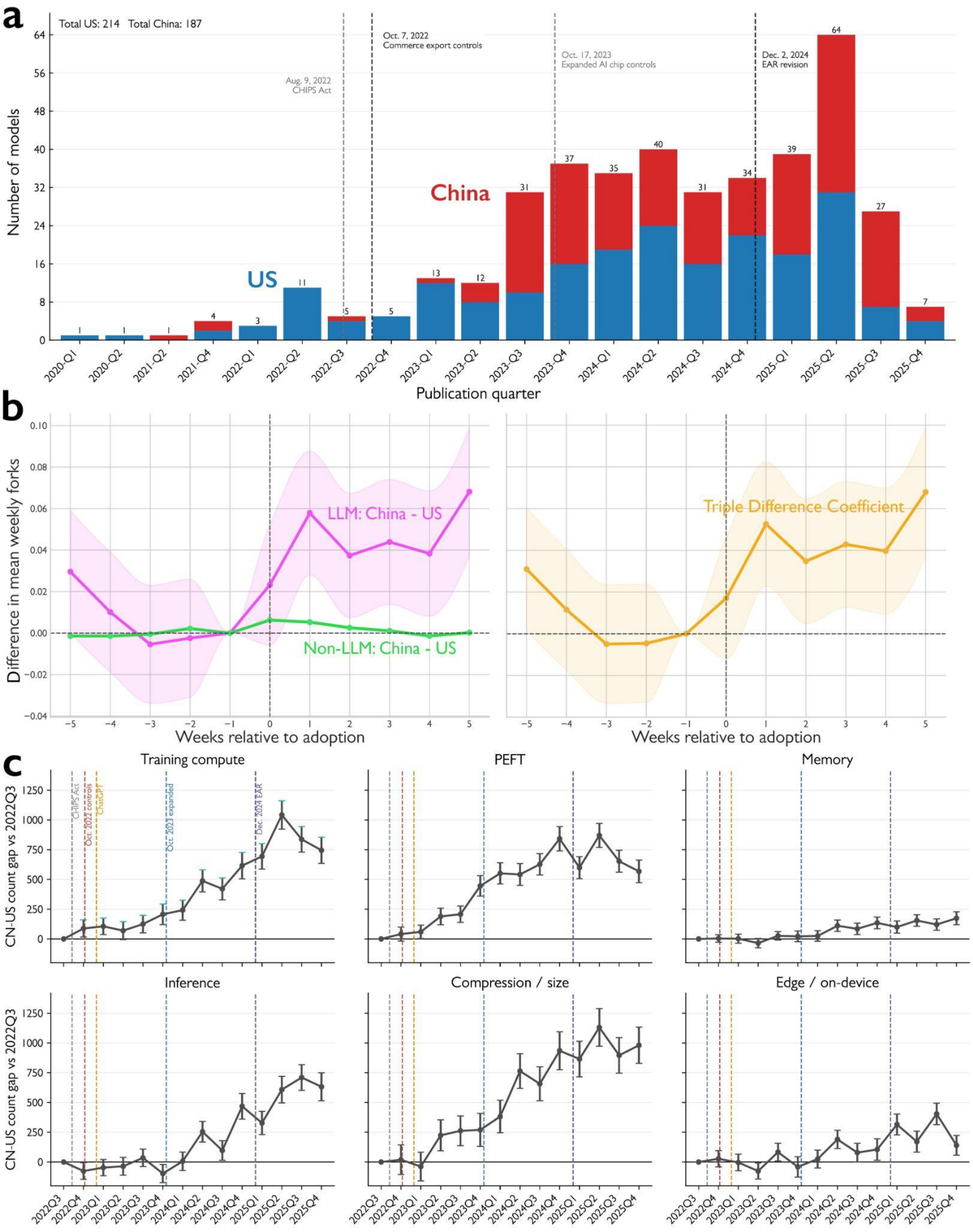

**Fig. 1. China's Open AI ecosystems around major U.S. policy events.** (A) U.S. and Chinese open-model releases over time. (B) Event-study estimates of repository forking around major U.S. policy events. The left panel reports China–U.S. difference-in-differences estimates separately for LLM and non-LLM repositories 5 weeks around a policy event. The right panel reports the triple-difference event-study estimates, defined as the difference between the LLM and non-LLM China–U.S. difference-in-differences estimates. All coefficients are normalized relative to one week before the policy event. (C) China–U.S. difference in the quarterly number of AI papers, with quarterly counts normalized to their 2020 Q3 baseline, for papers emphasizing compute restrictions, efficiency, memory, inference, compression, parameter-efficient fine-tuning (PEFT), and edge/on-device deployment.

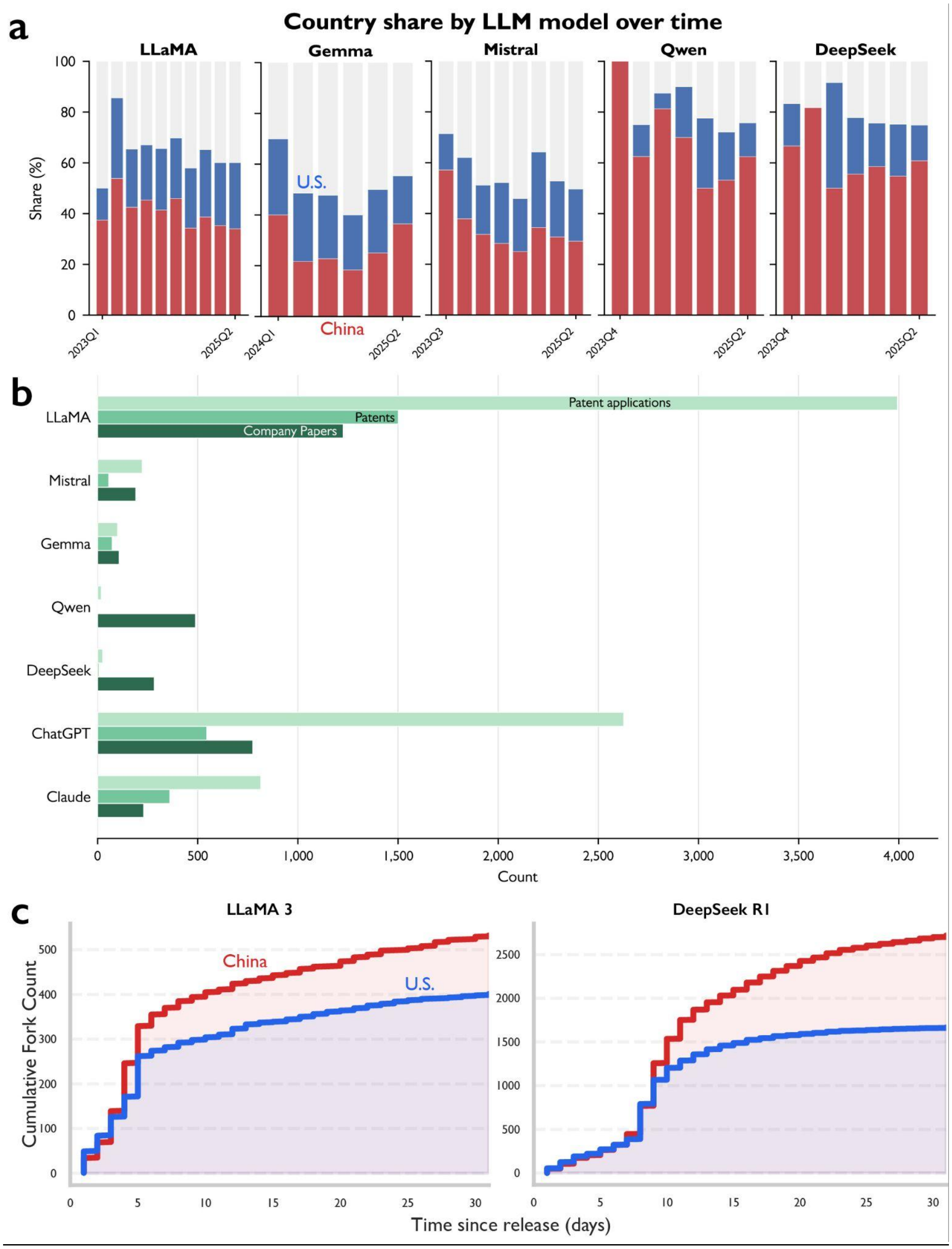

**Fig. 2. Cross-domain diffusion of open models.** (A) arXiv model mentions over time by author-country affiliation. (B) Model mentions in U.S. patent applications, granted patents, and arXiv papers. (C) Early open-source diffusion of DeepSeek-R1 and Llama 3, measured by cumulative forks after release.

**Acknowledgments:** We thank Zhangyizhe Yuan, Lu Tong, Can Hu, Yilin Chen, and Jiawen Zeng for excellent research assistance for this project.

**Funding:** N/A

**Author contributions:** W.J., N.K., B.L. contributed equally to this work and are listed as co-first authors in alphabetical order. T.S. and J.E. are the corresponding authors.

Conceptualization: W.J., N.K., B.L., T.S., J.E.

Methodology: W.J., N.K., B.L., T.S., J.E.

Investigation: W.J., N.K., B.L.

Visualization: W.J., N.K., B.L., J.E.

Project administration: W.J., N.K., B.L.

Supervision: T.S., J.E.

Writing – original draft: W.J., N.K., B.L.

Writing – review & editing: T.S., J.E.

**Competing interests:** J.E. maintains a commercial role with Google as a 25% visiting researcher. All other authors declare no competing interests.

**Data, code, and materials availability:** All data and code will be made available online.

**Supplementary Materials**

Materials and Methods

Figs. S1 to S9

# Supplementary Materials for

## U.S. Technological Containment and the Rise of China's Open AI Ecosystem

Wang Jin[1†], Nadav Kunievsky[2†], Bowen Lou[3†], Tianshu Sun[4*], James Evans[2,5*]

[1]Argyros College of Business and Economics, Chapman University; Orange, California, 92866, USA.
[2]Knowledge Lab, University of Chicago; Chicago, Illinois, 60637, USA.
[3]Marshall School of Business, University of Southern California; Los Angeles, California, 90089, USA.
[4]Cheung Kong Graduate School of Business; Beijing, 100006, China.
[5]Santa Fe Institute; Santa Fe, New Mexico, 87501, USA.

[†]These authors contributed equally to this work.
[*]Corresponding author. Email: tianshusun@ckgsb.edu.cn; jevans@uchicago.edu

**The PDF file includes:**

## Materials and Methods

Overview

Our empirical analyses combine six data sources: Chinese AI and open-source policy documents, a curated database of open or open-weight model releases, GitHub event records, arXiv metadata linked to author-country information, company-affiliation evidence for arXiv papers, and U.S. patent full-text records. The analyses trace observable patterns in policy orientation, model supply, developer adoption, scientific engagement, and patent disclosure.

GitHub event-study design

The GitHub event study measures repository forking around four U.S. policy events: the CHIPS and Science Act on 9 August 2022, the Commerce advanced-computing export controls on 7 October 2022, the expanded AI chip controls on 17 October 2023, and the 2 December 2024 Export Administration Regulations revision. For each event, repository activity is organized into an event-time panel indexed by repository $i$, event $e$, and relative week $t$, with $t=-1$ used as the omitted pre-event baseline. The outcome is the number of ForkEvent records for a repository-week. Repositories are classified as LLM-related if they appear in the generative-AI repository list; all other repositories in the panel form the non-LLM comparison group. Regional series in the GitHub analysis are based on event-hour timing. Events with hour less than or equal to 17 are assigned to the China-associated series, and later events are assigned to the U.S.-associated series. This is a time-zone proxy for regional activity. Fig. 1B plots one-percent-trimmed groups normalized to the week *-1* baseline. For group $g$ and relative week $t$, the plotted value is

$$\Delta_{gt} = mean(y_{gt}) - mean(y_{g,-1})$$

computed separately for China-associated LLM repositories, U.S.-associated LLM repositories, China-associated non-LLM repositories, and U.S.-associated non-LLM repositories. Confidence intervals are computed for the difference in group means between week $t$ and week *-1*. Standard errors are clustered by repository.

Efficiency and compute-restriction topics

The efficiency-research panel uses first-version arXiv papers from 2022 through the Dec 2025 metadata snapshot. The sample is restricted to AI-related categories: cs.AI, cs.CL, cs.CV, cs.LG, cs.NE, cs.RO, cs.IR, and stat.ML. Papers are linked to author-institution country rows. The unit of aggregation is an author-paper-country contribution, de-duplicated at the work, arXiv paper, author, and country level. A paper with both U.S. and Chinese institutional affiliations contributes to both country series. Papers are classified by title and abstract into six compute-restriction categories shown in Fig. 1C: compression or size reduction, parameter-efficient fine-tuning, inference efficiency, training-compute efficiency, memory efficiency, and edge or on-device deployment. A paper is included when its title or abstract directly links one of these topics to resource constraints such as compute, memory, model size, training cost, inference latency, or deployment requirements. Strong evidence includes an efficiency term in the title, a compression or parameter-efficient fine-tuning term, or a method term appearing near a resource term in the abstract.

The compression and size-reduction dictionary includes model compression, compressed model, compact model, compact network, lightweight model, lightweight network, lightweight neural network, lightweight transformer, small language model, tiny model, tiny network, model size reduction, fewer parameters, parameter reduction, quantization, quantized, low-bit, low precision, int4, int8, binary neural network, ternary neural network, pruning, pruned, sparsification, structured sparsity, unstructured sparsity, lottery ticket, weight sharing, knowledge distillation, distillation, and student-teacher.

The parameter-efficient fine-tuning dictionary includes parameter-efficient, parameter-efficient fine-tuning, PEFT, LoRA, low-rank adaptation, QLoRA, adapter tuning, adapter, adapters, prefix tuning, prompt tuning, P-tuning, BitFit, frozen language model, and trainable parameters.

The inference-efficiency dictionary includes inference efficiency, efficient inference, fast inference, inference acceleration, accelerating inference, low-latency, latency reduction, reduce latency, throughput, serving efficiency, efficient serving, real-time inference, decoding acceleration, fast decoding, speculative decoding, early exit, adaptive computation, dynamic inference, KV cache, and cache compression.

The training-compute dictionary includes training efficiency, efficient training, compute-efficient training, accelerating training, faster training, reduce training cost, training cost, GPU-hours, GPU-days, FLOPs, compute budget, compute efficient, computational efficiency, low compute, resource-efficient training, and communication-efficient training.

The memory-efficiency dictionary includes memory efficient, memory-efficient, memory footprint, reduce memory, memory reduction, VRAM, activation memory, optimizer memory, gradient checkpointing, activation checkpointing, memory-saving, out-of-memory, and low memory.

The edge/on-device dictionary includes edge AI, edge device, edge devices, mobile device, mobile devices, on-device, embedded device, embedded devices, resource-constrained, limited compute, limited memory, low-resource device, TinyML, microcontroller, IoT device, deployment on mobile, and deployment on edge.

For Fig. 1C, category counts are aggregated by country and quarter. For category $k$ and quarter $q$, the plotted estimand is

$$(CN_{kq} - US_{kq}) - (CN_{k2022Q3} - US_{k2022Q3})$$

where CN and US denote author-paper-country counts. Thus, the baseline quarter 2022Q3 is normalized to zero in every panel. The figure runs through 2025Q4. Confidence intervals use a Poisson count approximation for the China and U.S. counts in quarter q and in the 2022Q3 baseline.

### arXiv/OpenAlex model engagement

Scientific model engagement is measured by matching focal model-family names in arXiv titles and abstracts. The focal families are LLaMA, Mistral or Mixtral, Gemma, Qwen, and DeepSeek. The matching rules are case-insensitive regular expressions applied to title and abstract text, and duplicate paper-model pairs are removed. The resulting paper-model rows are dated by the arXiv first-version timestamp. For country attribution, arXiv papers are linked to author institutional countries; the country with the largest number of distinct authors is assigned to the paper, with countries grouped as United States, China, and other. Fig. 2A reports the country composition of model mentions over time by model family.

Patent model reference and visibility

Patent model reference is measured in U.S. patent applications and granted-patent full-text records from the United States Patent and Trademark Office (USPTO). XML records are parsed into structured patent records, and patent descriptions are searched for model terms from a local LLM-term dictionary. For each patent, the analysis records the patent identifier, matched term, term frequency, production date, and publication date, then removes duplicate patent-term rows. Counts are based on unique patent identifiers mentioning each model term after the corresponding model release date. Fig. 2B compares these patent counts with arXiv model-mention counts for the same selected model families.

Company-affiliated arXiv model engagement

Company-affiliated model engagement is measured among arXiv papers that mention selected model families in titles or abstracts after each model's release date. The company universe is constructed from active company records linked to U.S. or Chinese organizations and restricted to companies appearing on at least five cs.AI arXiv papers. Where local office location differs from parent-company origin, the plotted country assignment uses parent-company origin. The final company universe contains 46 parent-origin company groups, 35 assigned to the United States and 11 to China.

Company evidence is identified from three sources: paper-specific affiliation strings in OpenAlex work records, paper-specific arXiv source snippets when available, and OpenAlex author-profile affiliations in the exact paper year. The plotted rule counts a paper-model-country observation if any company assigned to that country appears under either the paper-specific evidence rule or the exact-year author-profile rule. A paper can count once in both the U.S. and Chinese company series if it contains qualifying company evidence for both countries.

Open-model repository diffusion

Repository diffusion is measured from GitHub events for canonical repositories associated with DeepSeek-R1, Qwen 2.5, Qwen 3, Gemma, LLaMA, and Mistral. The diffusion figures use ForkEvent records because forking is an observable act of copying a repository into a developer-controlled namespace. Events are deduplicated by event type, actor, and repository so that a given actor contributes at most once to a repository's fork curve. The cumulative diffusion curve for each country is constructed by aggregating all repository forks observed up to each day since the model's release date across the set of canonical repositories. Main-text panels show 31-day curves for DeepSeek-R1 and Llama 3; supplementary panels show 31-day and 361-day curves for all tracked repositories. As in the GitHub event study, event hour is used as a regional proxy: hour less than or equal to 17 for the China-associated series and hour greater than 17 for the U.S.-associated series.

## Open-Source Performance Boost from Hugging Face Leaderboard

The size-bucket frontier figures Fig. S3-Fig. S5 use model-level results from the Hugging Face Open LLM Leaderboard v2, specifically the open-llm-leaderboard/contents dataset on Hugging Face and associated Hugging Face Hub model metadata, downloaded and processed on June 5, 2026. The final analysis keeps 1,981 U.S.- and China-origin open-weight models: 189 official releases and 1,792 community fine-tuned or derived models. Country of origin is assigned at the model-family level: official releases are linked to source organizations and families, such as Meta/Llama for the U.S. or Qwen/Alibaba for China, and non-official models inherit the origin country of the detected base family when the lineage is unambiguous. Release month comes from the leaderboard's Upload To Hub Date; when that field is missing, we use the leaderboard Submission Date as a fallback. The plotted window runs from April 2023 through March 2025. Models are split by reported parameter count into small models, <10B, medium models, 10-30B, and large models, >=30B; each line shows the best benchmark score observed up to each month within a country, benchmark, and size bucket.

Fig. S6 compares two similar-size official U.S./China model pairs: Llama-3.2-3B-Instruct vs. Qwen2.5-3B-Instruct, and Meta-Llama-3-8B-Instruct vs. Qwen2-7B-Instruct. Follow-up models are linked to an official family using Hugging Face metadata and model identifiers: a model is treated as a follow-up when it declares a base model, has a family-base link, or is classified as a probable downstream tuned model from naming and lineage evidence. For the 3B pair, this gives 55 Llama follow-ups and 207 Qwen follow-ups; for the 7B/8B pair, this gives 287 Llama follow-ups and 33 Qwen follow-ups. Individual dots are follow-up benchmark scores, large circles are official model scores, and horizontal bars mark the best follow-up score within each country-benchmark group; official zero scores are treated as missing and omitted.

The benchmarks are the six component tasks reported by the Open LLM Leaderboard v2. IFEval measures instruction-following accuracy. BBH, or Big-Bench Hard, measures performance on difficult reasoning tasks. MATH Lvl 5 measures accuracy on hard competition-style math problems. GPQA is Graduate-Level Google-Proof Q&A, a difficult science and reasoning benchmark designed to be resistant to simple web lookup. MuSR measures multi-step soft reasoning over longer contexts. MMLU-Pro is a harder version of MMLU covering broad professional and academic knowledge with more challenging answer choices.

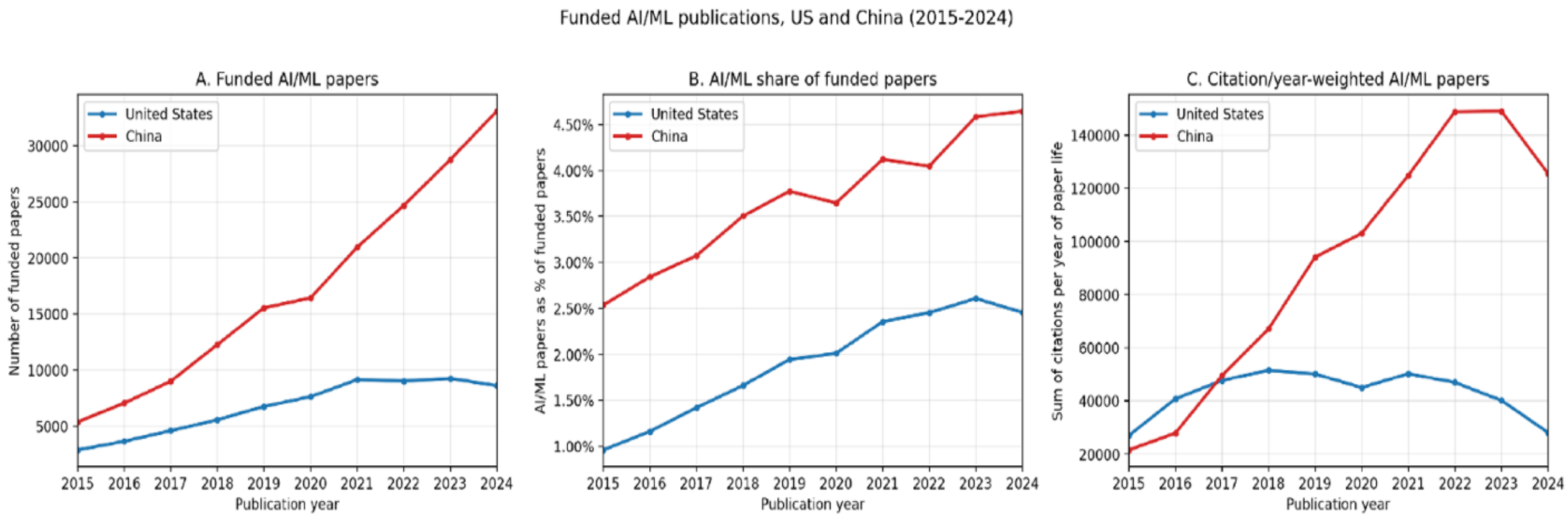

Fig. S1. Funded AI/ML publication trends for the United States and China.

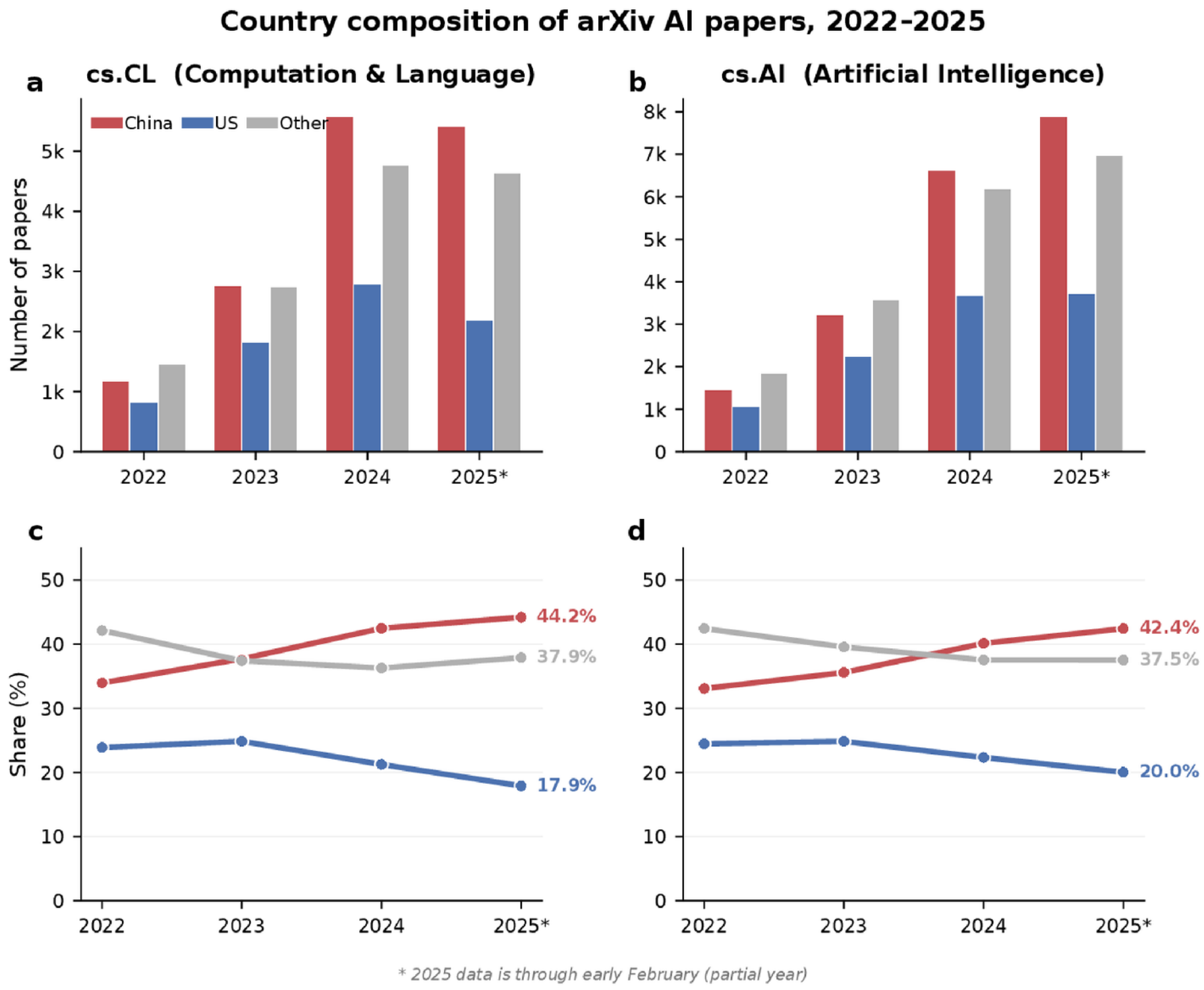

Fig. S2. Country composition of arXiv AI categories.

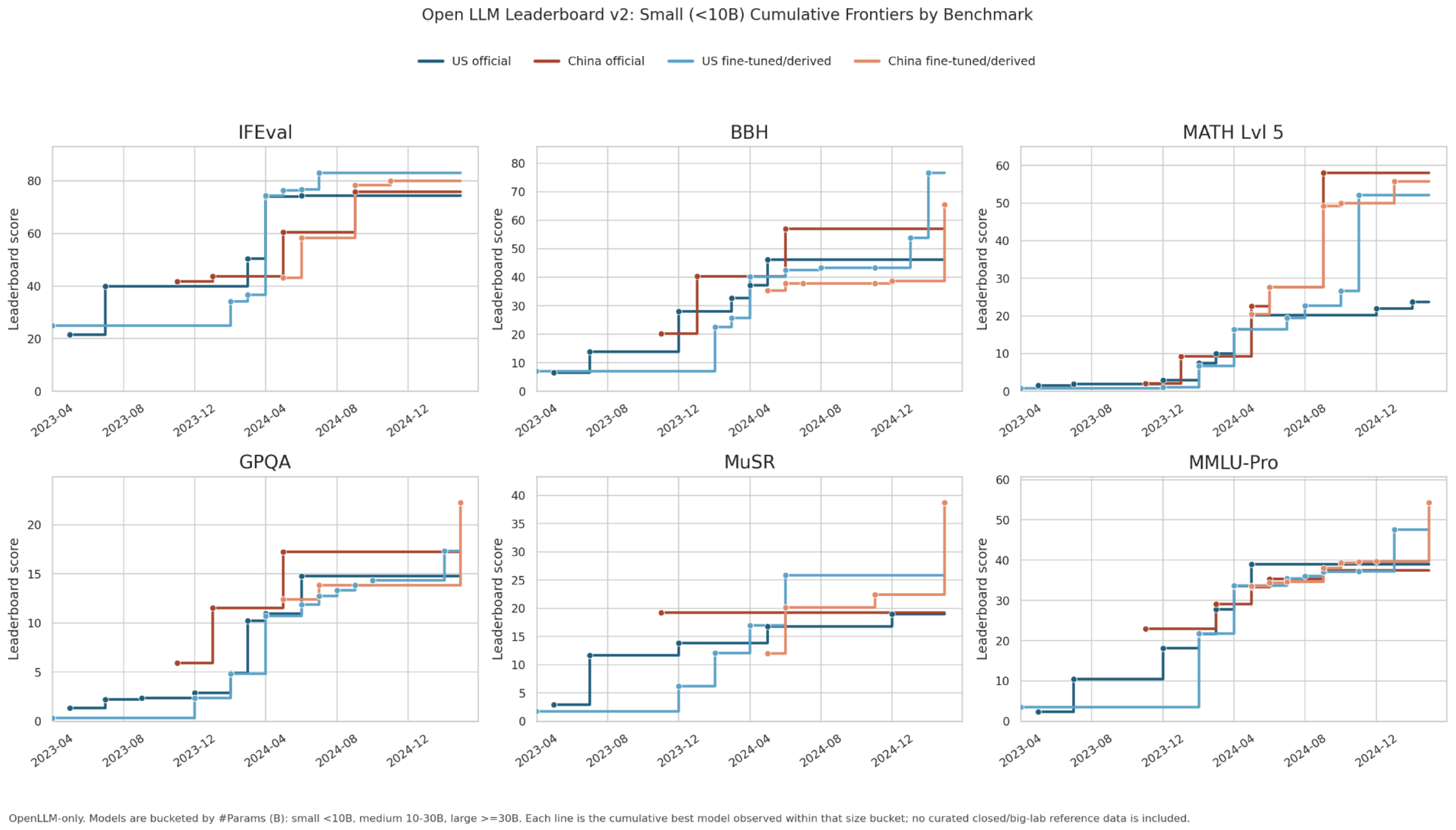

Fig. S3. Comparison of top-performing official Chinese and U.S. open models on the Hugging Face Open LLM Leaderboard and their descendant models, restricted to models with fewer than 10 billion parameters.

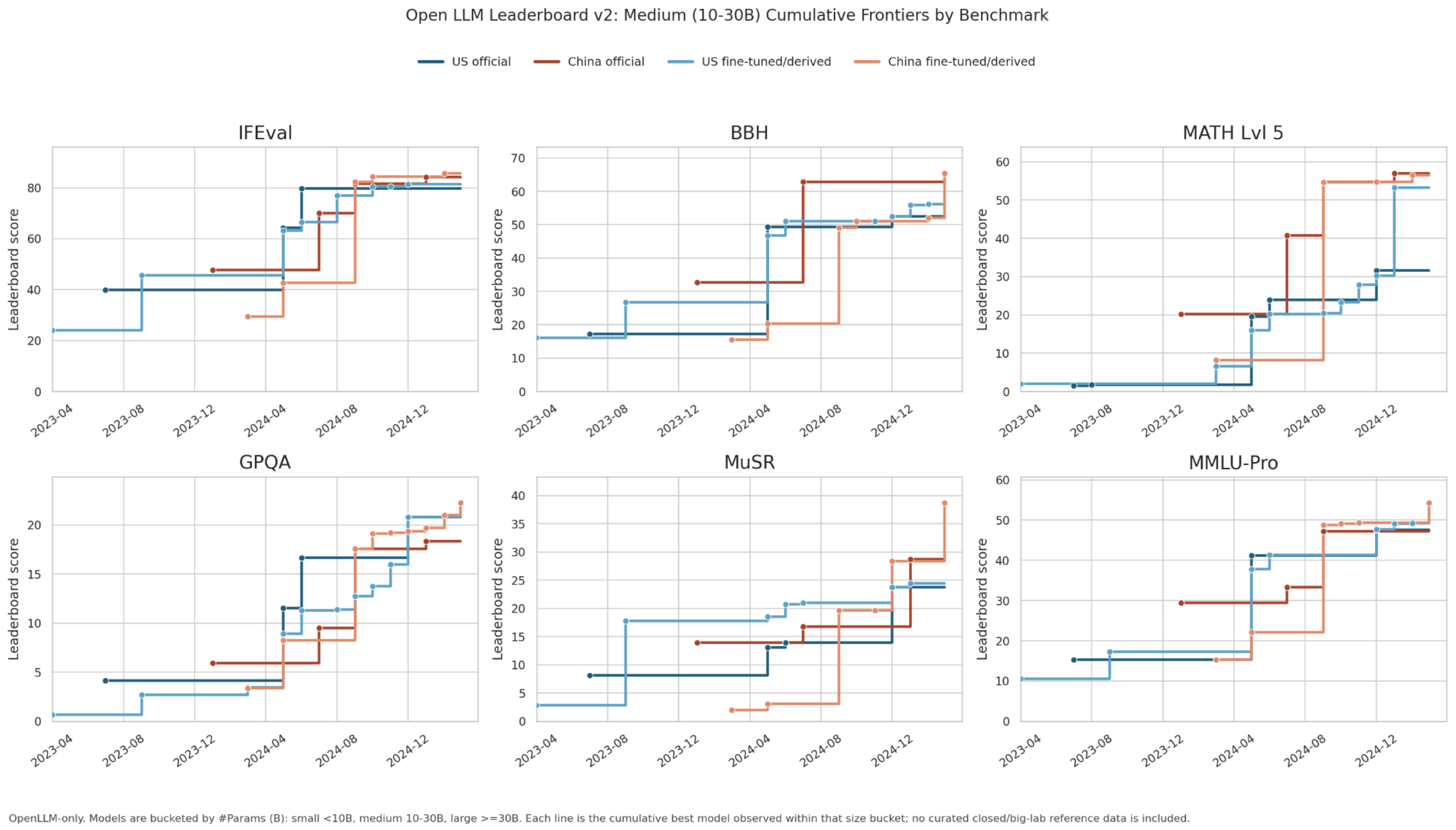

Fig. S4. Comparison of top-performing official Chinese and U.S. open models on the Hugging Face Open LLM Leaderboard and their descendant models, restricted to models with 10 billion and 30 billion parameters.

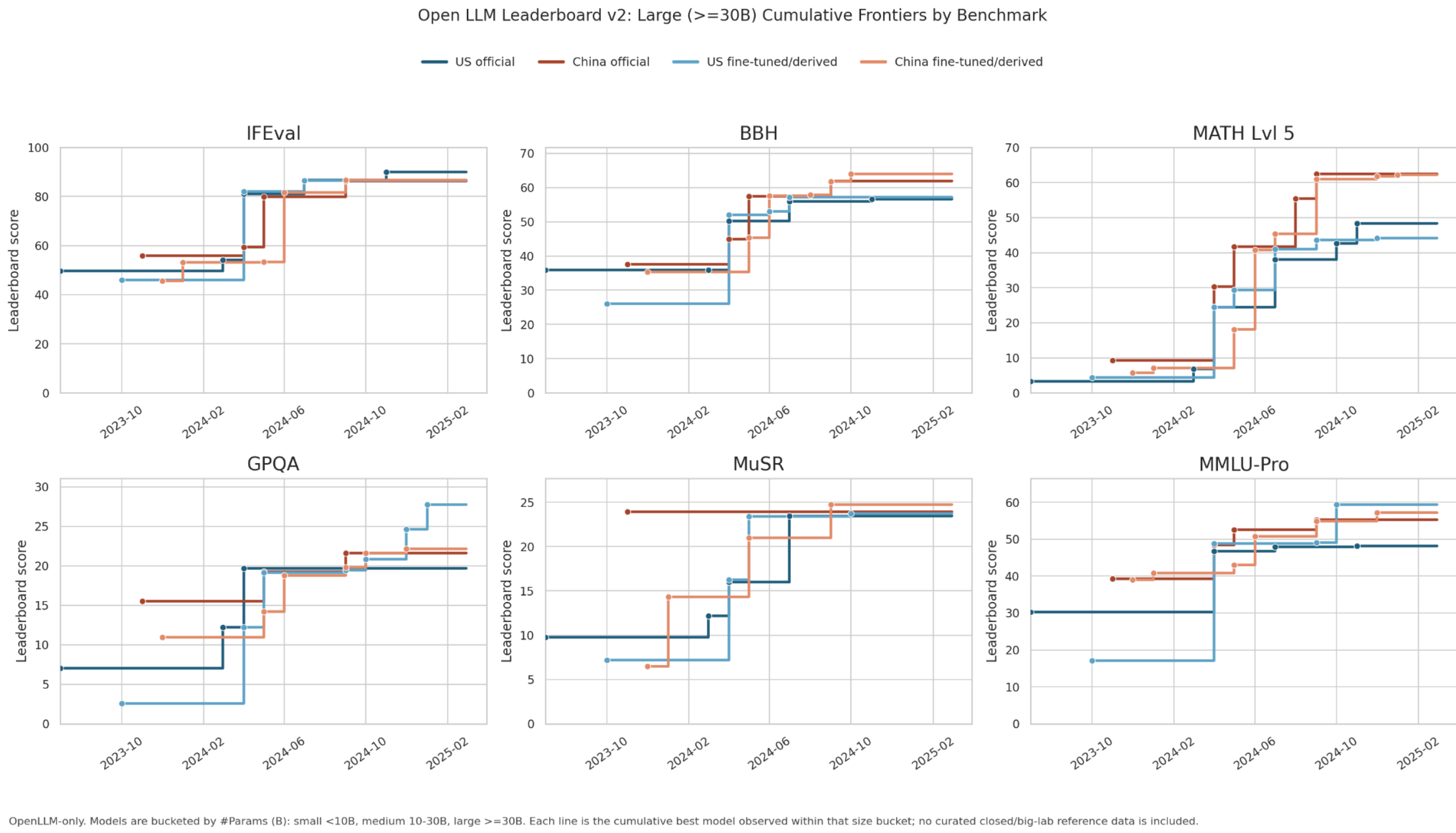

Fig. S5. Comparison of top-performing official Chinese and U.S. open models on the Hugging Face Open LLM Leaderboard and their descendant models, restricted to models with more than 30 billion parameters.

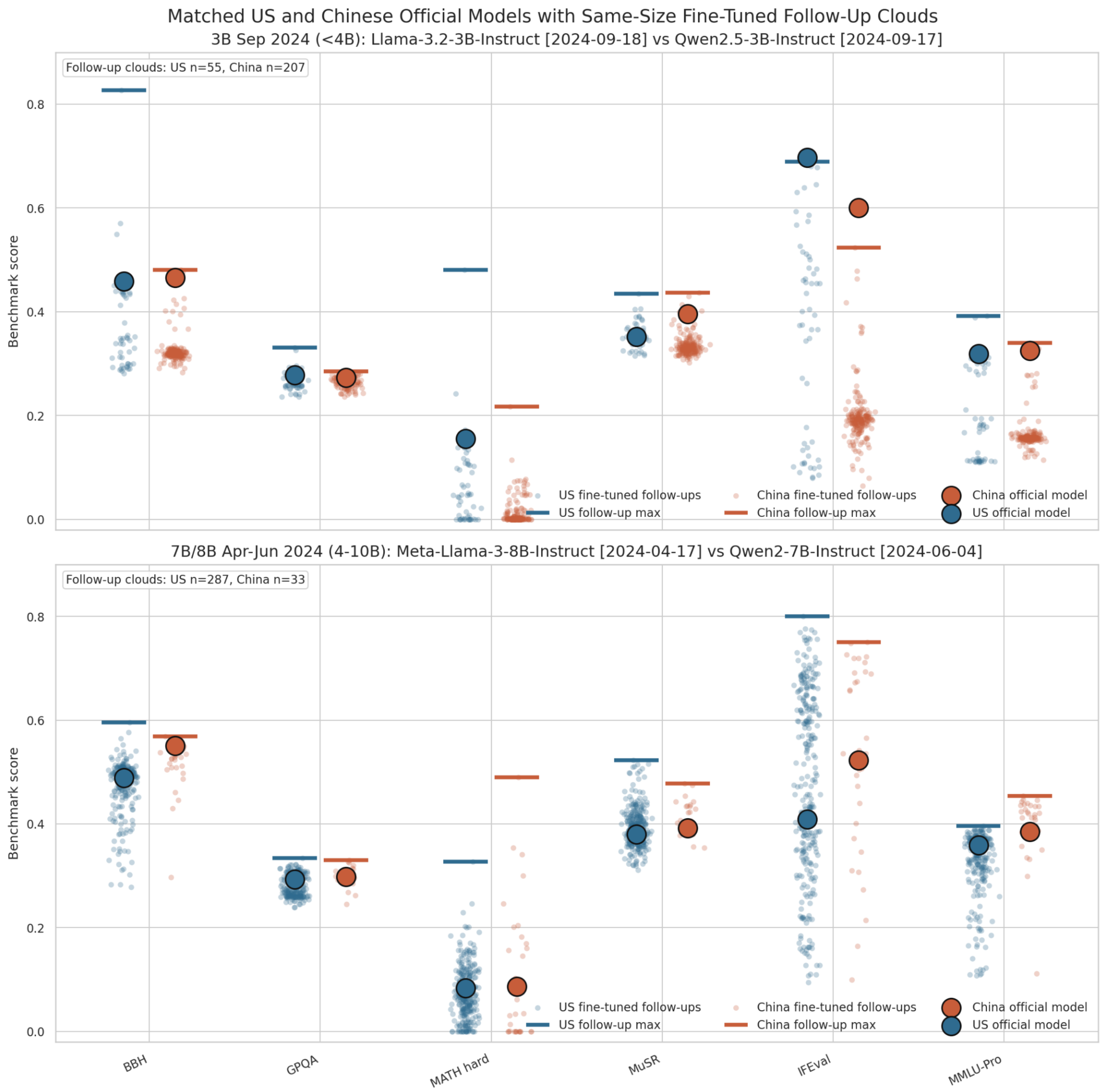

Fig. S6. Comparison of benchmark scores for similarly sized official Llama and Qwen model releases with the distribution of benchmark performances achieved by their descendant models.

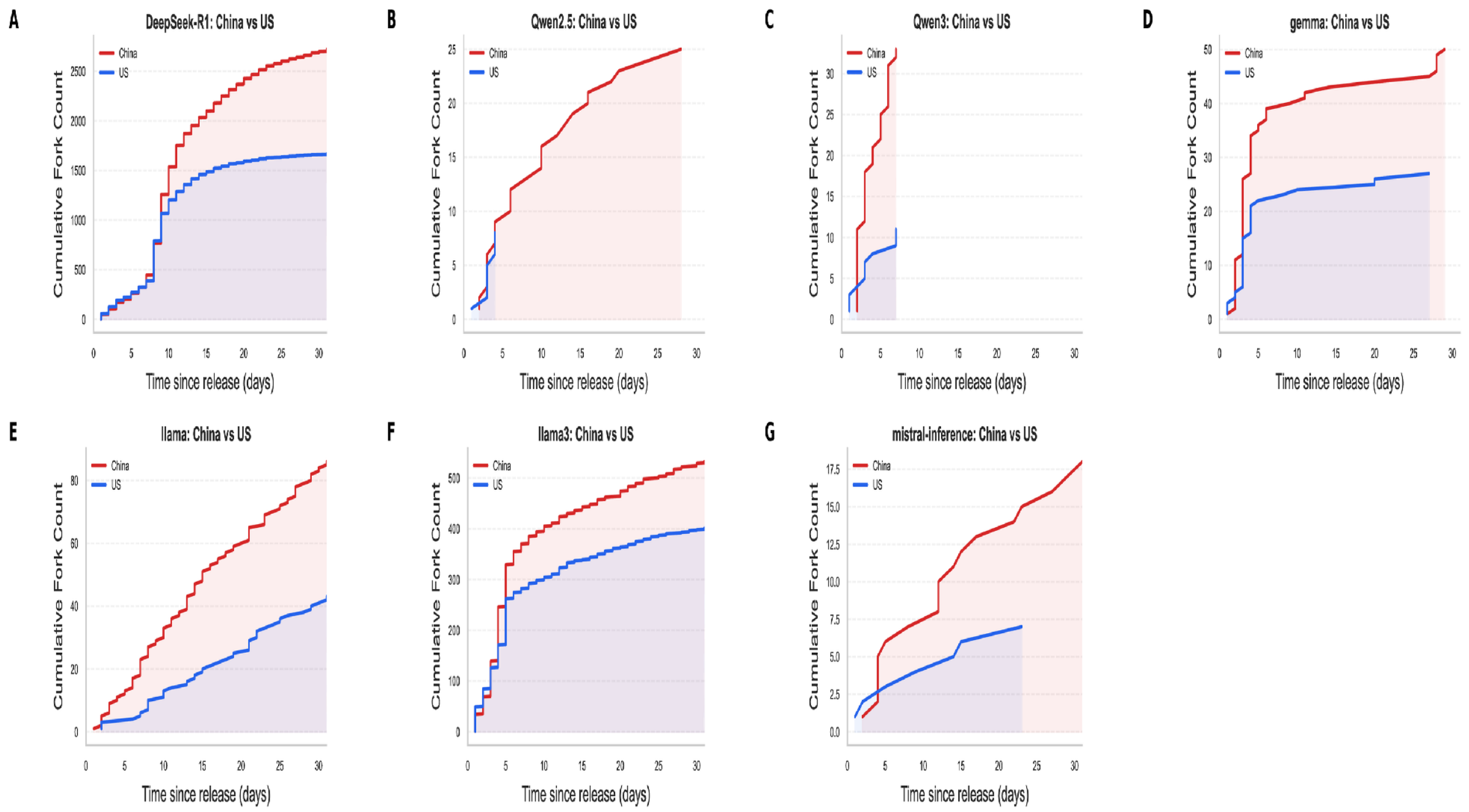

Fig. S7. Thirty-one-day open-source diffusion profiles for all tracked model repositories.

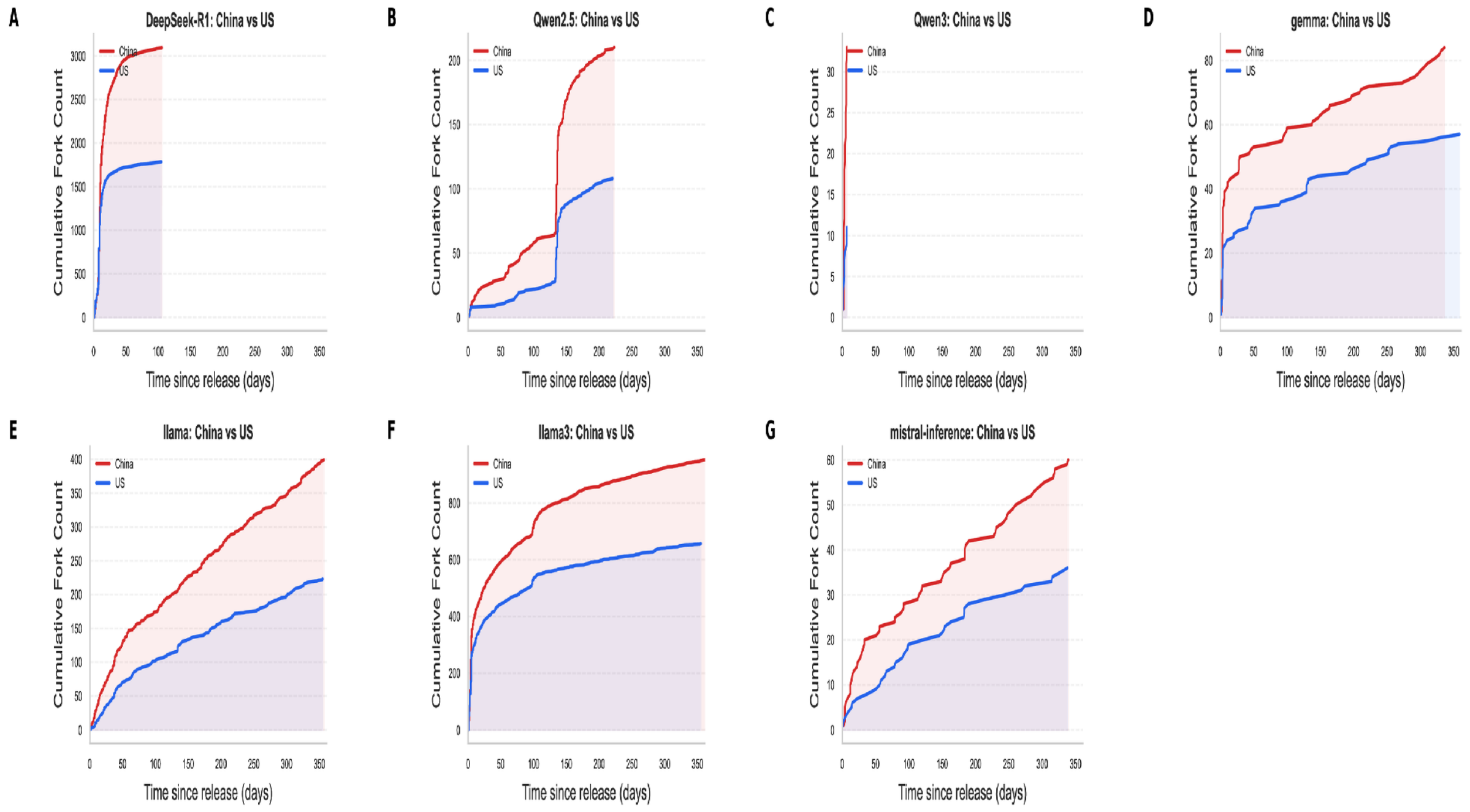

Fig. S8. Three-hundred-sixty-one-day open-source diffusion profiles for all tracked model repositories.

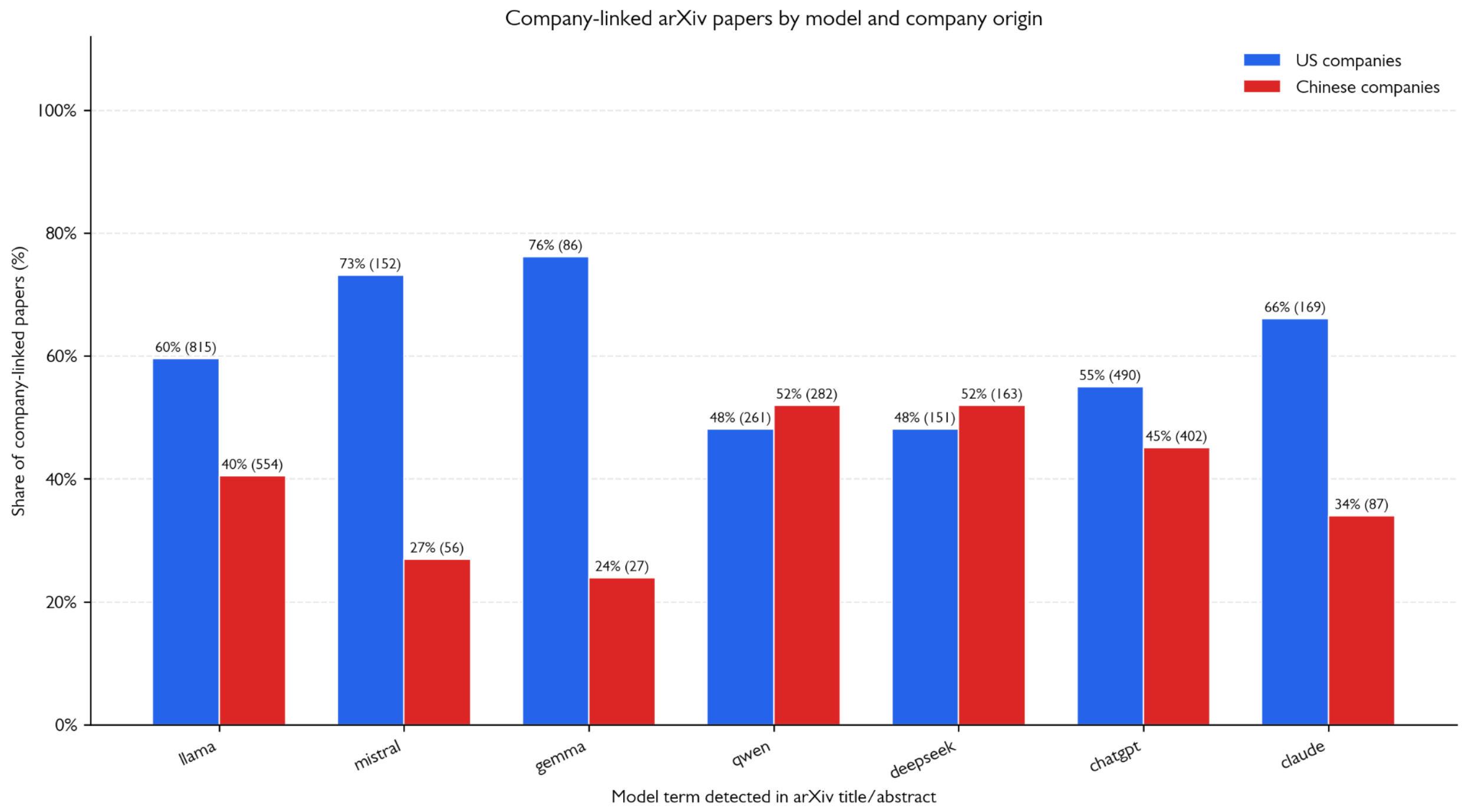

Fig. S9. Company-affiliated arXiv model engagement.